\documentclass[12pt]{article}
\usepackage[utf8]{inputenc}
\usepackage{authblk}
\usepackage{setspace}
\usepackage[margin=1.25in]{geometry}
\usepackage{graphicx}
\usepackage{subcaption}
\usepackage{amsmath}
\usepackage{lineno}
\usepackage{lipsum}
\usepackage[comma,square,numbers,sort&compress]{natbib}

\usepackage[dvipsnames]{xcolor}
\usepackage{hyperref}
\hypersetup{                                                          
pdftoolbar    = true,            
pdfmenubar    = true,            
colorlinks    = true,                                                          
linkcolor     = blue,                                                          
citecolor     = blue,                                                       
linktocpage   = true,  
linktocpage   = true,
urlcolor      = CadetBlue,   
}
\begin{document}
\title{Local multiplicity fluctuations in Pb$-$Pb collisions at $\sqrt{s_{\rm{NN}}}$ = 2.76 TeV with ALICE at the LHC}
\author[$ $]{Sheetal Sharma$^{*}$ and Ramni Gupta$^{\dagger}$ {(for the ALICE Collaboration)}}

\affil[$ *\dagger$]{Department of Physics, University of Jammu, Jammu, India}
\affil[$ $]{$^{*}$sheetal.sharma@cern.ch}

\onehalfspacing
\maketitle

\date{}

\begin{abstract}
Local multiplicity fluctuations are an useful tool to understand the dynamics of the particle production and the phase-space changes from quarks to hadrons in ultrarelativistic heavy-ion collisions. The study of scaling behavior of multiplicity fluctuations in geometrical configurations in multiparticle production can be performed using the factorial moments and recognized in terms of a phenomenon referred to as intermittency. 

In this contribution, the analysis of the factorial moment is presented for the multiplicity distributions of charged particles produced in Pb$-$Pb collisions at $\sqrt{s_{\rm{NN}}}$ = 2.76 TeV, recorded with the ALICE detector at the LHC. The normalized factorial moments (NFM), $F_{q}$ of the spatial configurations of charged particles in two-dimensional angular ($\eta,\varphi$) phase space are calculated. For a system with dynamic fluctuations due to the characteristic critical behavior near the phase transition, $F_{q}$ exhibits power-law growth with increasing bin number or decreasing bin size which indicates self-similar fluctuations. Relating the $q^{\rm{th}}$ order NFM ($F_{q}$) to the second-order NFM ($F_{2}$), the value of the scaling exponent ($\nu$) is extracted,  which indicates the order of the phase transition within the framework of Ginzburg-Landau theory. The dependence of scaling exponent on the $p_{\rm{T}}$ bin width will be presented. The measurements are also compared with the corresponding results from the AMPT model and a Toy Monte Carlo (MC) simulation.


\end{abstract}

\newpage
\section{Introduction and Analysis Details}
Heavy-ion collisions offer a unique opportunity to investigate the creation and properties of quark--gluon plasma (QGP). As the QGP rapidly cools down, it undergoes a phase transition into hadronic matter that provides crucial information about the properties of the system. Spatial configurations of the charged particles produced in two dimensional ($\eta, \varphi$) phase space are proposed to be studied to understand the multiparticle production at LHC \cite{Hwa:2011bu}. The scaling behavior of the NFM of the multiplicity distributions is studied in the framework of the intermittency methodology \cite{Bialas:1985jb,Hwa:1992uq, Hwa:1992cn, Hwa:2011bu}.

Event-by-event intermittency analysis is performed for charged particles produced in the midrapidity region and full azimuth in Pb--Pb collisions at $\sqrt{s_{\rm{NN}}}$ = 2.76 TeV recorded by the ALICE detector. The $F_{\rm{q}}$ coefficients are determined using the methodology described in Ref.~\cite{Gupta:2019zox} using Eq.~1 from Ref.~\cite{Sharma:2021cyp} 
for different number of bins M (partitioning in ($\eta, \varphi$) phase space). For the spatial fluctuations in the data that are scale independent, $F_{\rm{q}}(M)$ is proportional to $M^{\phi_{\rm{q}}}$ for \textit{q} $\geq 2$ and for $\phi_{\rm{q}} > 0$, is the intermittency index. This power law dependence is defined as intermittency. Independent of the observation of this behavior, a scaling exponent ($\nu$) is obtained from the dependence of $F_{\rm{q}}(M)$ (with \textit{q} $\geq 3$) on $F_{\rm{2}}(M)$. This scaling exponent is 1.304 in the framework of Ginzburg-Landau (GL) theory for the second order phase transition \cite{Hwa:1992cn}.

\begin{figure}[htp]
\centering
  \includegraphics[width=0.39\textwidth]{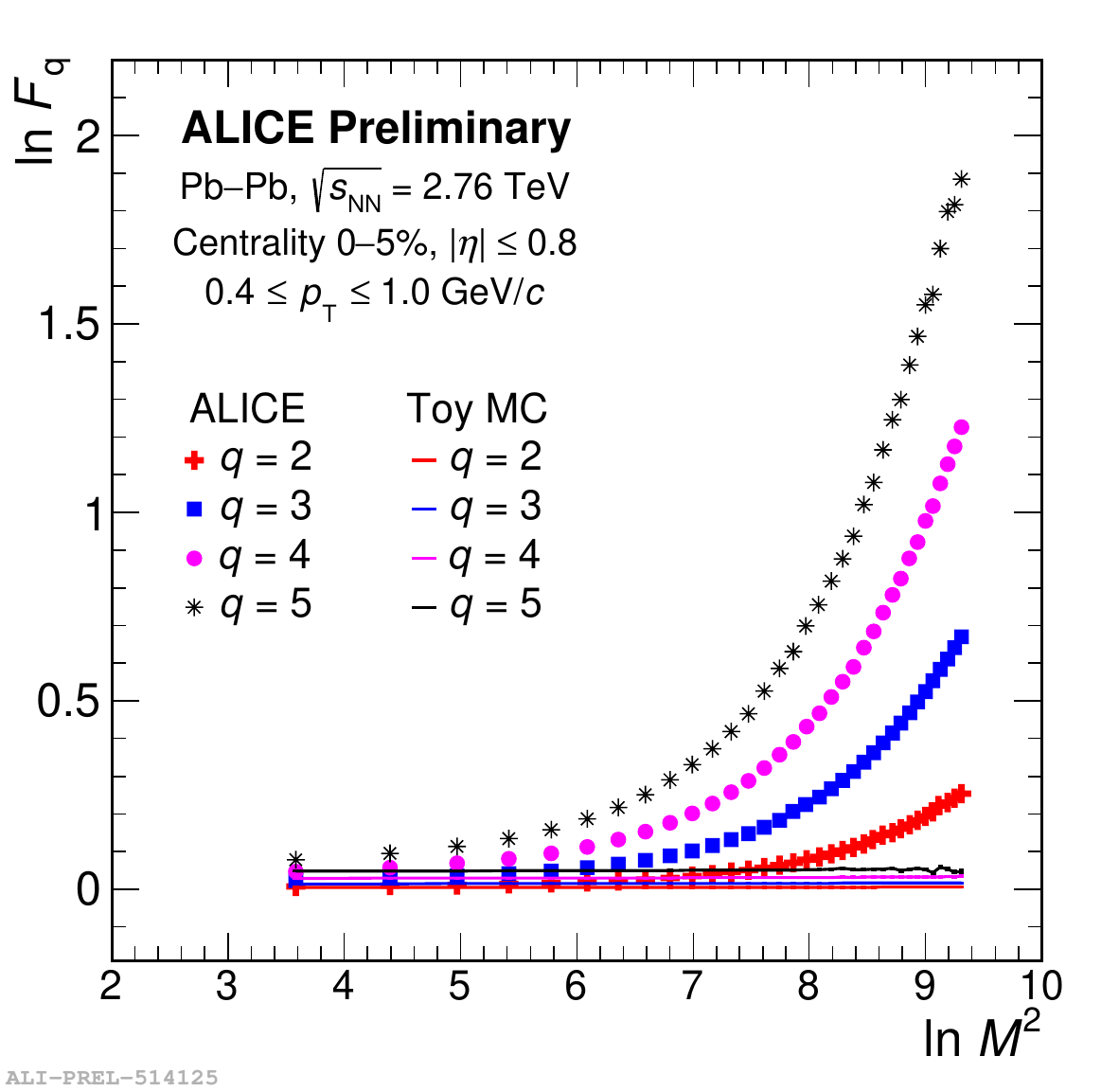}
  \includegraphics[width=0.58\textwidth]{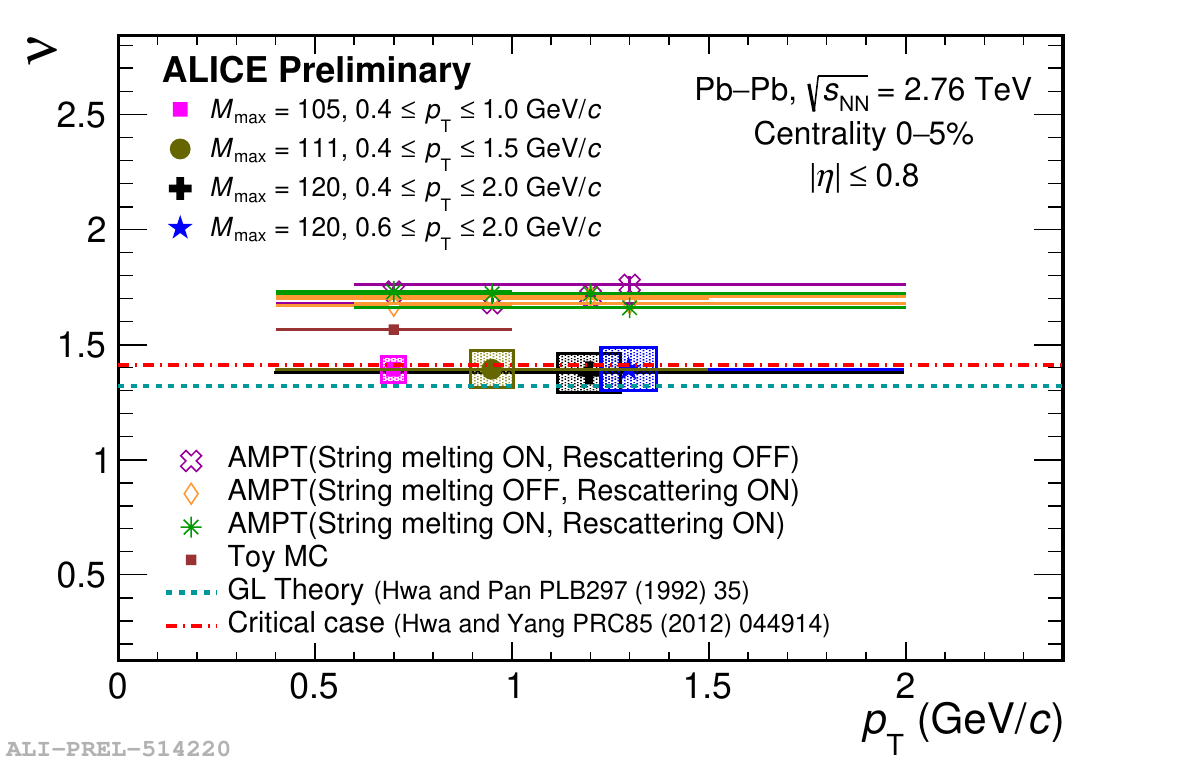}
  \caption{(Left) Normalized factorial moments up to $5^{\rm{th}}$ order as a function of $M^{2}$ compared with calculations from a Toy Monte Carlo (MC) simulation; (Right) Transverse momentum dependence of the scaling exponent compared with expectations from the AMPT model, Toy MC simulation, and GL theory. }
  \label{mscalnnu}
\end{figure}

\section{Results and Conclusion}
For ALICE data, intermittency analysis is performed for the charged particles produced in $|\eta| \leq 0.8$ and full azimuth in different $p_{\rm{T}}$ bins. A Toy MC simulation with only statistical fluctuations is performed as baseline. The $M^{\rm{2}}$ dependence of $F_{\rm{q}}$ ($q$ = 2, 3, 4, 5) shows a significant deviation from baseline as presented in the left panel of Fig. \ref{mscalnnu}. 
The scaling exponent for different $p_{\rm{T}}$ bin widths is presented in the right panel of Fig. \ref{mscalnnu}. The data is also compared with AMPT model calculations, as well as GL theory predictions.

 In summary, an intermittency signal, i.e., a linear behavior between ln~$F_{\rm{q}}$ and ln~$M^{\rm{2}}$ is observed at larger M values, indicating a scale-invariant pattern in the distribution of the particles. The scaling exponent shows no dependence on the $p_{\rm{T}}$ bin width within the uncertainties and is consistent with models that include critical fluctuations within the experimental uncertainties.


\begin{thebibliography}{50}
\bibitem{Hwa:2011bu}
R.~C.~Hwa and C.~B.~Yang,
Phys. Rev. C \textbf{85} (2012), 044914
doi:10.1103/PhysRevC.85.044914
[arXiv:1111.6651 [nucl-th]].

\bibitem{Bialas:1985jb}
A.~Bialas and R.~B.~Peschanski,
Nucl. Phys. B \textbf{273} (1986), 703-718
doi:10.1016/0550-3213(86)90386-X

\bibitem{Hwa:1992uq}
R.~C.~Hwa and M.~T.~Nazirov,
Phys. Rev. Lett. \textbf{69} (1992), 741-744
doi:10.1103/PhysRevLett.69.741

\bibitem{Hwa:1992cn}
R.~C.~Hwa and J.~c.~Pan,
Phys. Lett. B \textbf{297} (1992), 35-38
doi:10.1016/0370-2693(92)91065-H

\bibitem{Gupta:2019zox}
R.~Gupta and S.~K.~Malik,
Adv. High Energy Phys. \textbf{2020} (2020), 5073042
[erratum: Adv. High Energy Phys. \textbf{2020} (2020), 7319894]
doi:10.1155/2020/5073042
[arXiv:1911.13111 [hep-ex]].

\bibitem{Sharma:2021cyp}
S.~Sharma and R.~Gupta,
SciPost Phys. Proc. \textbf{10} (2022), 024
doi:10.21468/SciPostPhysProc.10.024
[arXiv:2110.11901 [hep-ph]].
\end{thebibliography}
\end{document}